\documentstyle[12pt,thmsa]{article}

\begin{document}

\begin{titlepage}
\begin{center}
{\Large \bf Instability of the Randall-Sundrum Model and Exact Bulk Solutions}

\vspace{5mm}

\end{center}

\vspace{5 mm}

\begin{center}
{\bf Hongya Liu
\footnote[1]{ Department of Physics, Dalian University of Technology,
Dalian, 116024, People's  Republic of China.}
$^{,}$\footnote[2]{Corresponding author. Email: hyliu@dlut.edu.cn}
and Guowen Peng \footnotemark[1]}

\vspace{3mm}

\end{center}

\vspace{1cm}

\begin{center}
{\bf Abstract}
\end{center}
\baselineskip 24pt
Five dimensional geodesic equation is used to study the gravitational
force acted on a test particle in the bulk of the Randall-Sundrum two-brane model.
This force could be interpreted as the gravitational attraction from
matters on the two branes and may cause the model to be unstable.
By analogy with star models in astrophysics, a fluid RS model is proposed in which
the bulk is filled with a fluid and this fluid has an anisotropic pressure to balance the
gravity from the two branes. Thus a class of exact bulk solutions is obtained
which shows that any 4D Einstein solution with a perfect fluid source can be
embedded in $y=$ constant hypersurfaces in the bulk to form an equilibrium
state of the brane model. By requiring a 4D effective curvature to have a
minimum, the compactification size of the extra dimension is discussed.

\vspace{1cm}
{\noindent \bf KEY  WORDS:} Higher dimensions, Brane models.

\end{titlepage}
\baselineskip=24pt

\section{INTRODUCTION}

There is a strong interest in the possibility that our universe is a 3-brane
embedded in a higher dimensional space. It has been proposed that the large
hierarchy between the weak scale and the fundamental scale of gravity can be
eliminated if the extra compact dimensions are large\cite{ADD}. An
alternative solution to the hierarchy problem, proposed by Randall and
Sundrum (RS), assumes that our universe is a negative tension brane
separated from a positive tension brane by a five-dimensional anti-de Sitter
($AdS_5$) bulk space\cite{RS1}. This does not require a large extra
dimension: the hierarchy problem is solved by the special properties of the
AdS space. A similar scenario to the RS one is that of Horava and Witten\cite%
{HW}, which arises within the context of M-theory.

The RS two-brane solution satisfies the 5D Einstein equations 
\begin{equation}
R_{AB}-\frac 12g_{AB}R=-\frac 1{4M^3}\left\{ \Lambda g_{AB}+\left[ \lambda
_{vis}\delta (y-y_c)+\lambda _{hid}\delta (y)\right] g_{\mu \nu }\delta
_A^\mu \delta _B^\nu \right\} \;  \label{RSeq}
\end{equation}
with a non-factorizable 5D metric being 
\begin{equation}
ds^2=W^2(y)\widetilde{\eta }_{\alpha \beta }dx^\alpha dx^\beta +dy^2\;.
\label{RSmetr}
\end{equation}
Here and in the following we use signature ($-++++$), and we use upper case
Latin letters to denote 5D indices ($0,1,2,3,5$) and lower case Greek
letters to denote 4D indices ($0,1,2,3$). In (\ref{RSeq}) and (\ref{RSmetr}%
), the ``warp'' factor $W(y)$ is 
\begin{equation}
W(y)=e^{-k\left| y\right| }\;,  \label{W(y)}
\end{equation}
and $\lambda _{vis}$, $\lambda _{hid}$ and $\Lambda $ are 
\begin{equation}
\lambda _{hid}=-\lambda _{vis}=24M^3k,\;\;\Lambda =-24M^3k^2\;.
\label{lambda}
\end{equation}
In this solution, the fifth dimension has the $Z_2$ reflection symmetry $%
(x,y)\rightarrow (x,-y)$ with $-y_c\leq y\leq y_c$. The hidden brane and the
visible brane are located at $y=0$ and $y=y_c$, respectively.

The instability of the RS model has received extensive studies [4, 5]. In
this paper, we wish to approach this subject from a different perspective.
The paper is arranged as follows. In section 2 we use the 5D geodesic
equations to study the instability of the model. In section 3 we introduce a
5D anisotropic fluid in the bulk and derive a hydrostatic equilibrium
equation of the bulk fluid along the $y$-direction. In section 4 we look for
exact solutions of the 5D Einstein equations. In section 5 we discuss the
embedding of several well known 4D exact solutions. In section 6 we study
the compactification size of the fifth dimension.

\section{GRAVITATIONAL FORCE IN THE BULK}

In this section we study the gravitational interaction between matters on
the two branes of the RS model. It is known from the brane-world scenario
that Standard Model (SM) particles are confined to branes while gravitons
can freely propagate in the bulk. Now let us consider a test particle in the
bulk. It is reasonable to expect that the motion of a bulk test particle,
which is acted on by the gravitational force only, is described by the
following 5D geodesics\cite{Youm}: 
\begin{equation}
\frac{d^2x^A}{d\tau ^2}+\Gamma _{BC}^A\frac{dx^B}{d\tau }\frac{dx^C}{d\tau }%
=0\;,  \label{geoEq}
\end{equation}
where $\Gamma _{BC}^A$ is the Christoffel symbol for the 5D metric $g_{AB}$
and $d\tau ^2=-ds^2=-g_{AB}dx^Adx^A$. It is known that 5D geodesic equations
(\ref{geoEq}) may yield extra 4D forces\cite{Youm}\cite{MWL}. In this paper,
we are not going to study this kind of extra forces; we only wish to study
particle's motion along the fifth direction. From (\ref{geoEq}), the 5D
gravitational force can be defined as 
\begin{equation}
F^A=-\Gamma _{BC}^A\frac{dx^B}{d\tau }\frac{dx^C}{d\tau }\;.  \label{FA}
\end{equation}
Using (\ref{RSmetr}) and (\ref{W(y)}) we find that the fifth component of $%
F^A$ is 
\begin{equation}
\frac{d^2y}{d\tau ^2}=F^5=-\Gamma _{BC}^5\frac{dx^B}{d\tau }\frac{dx^C}{%
d\tau }=\varepsilon k\left[ 1+\left( \frac{dy}{d\tau }\right) ^2\right] \;,
\label{A=5a}
\end{equation}
where 
\begin{equation}
\varepsilon =\left\{ 
\begin{array}{l}
1\quad \;\,for\quad y>0 \\ 
-1\quad for\quad y<0%
\end{array}
\right. \;.  \label{A=5b}
\end{equation}
So we find 
\begin{eqnarray}
F^5 &>&0\quad for\quad y>0\;,  \nonumber \\
F^5 &<&0\quad for\quad y<0\;.  \label{A=5c}
\end{eqnarray}
This result shows that the force $F^5$ acting on the bulk test particle
points from the hidden brane at $y=0$ to the visible brane in both $y>0$ and 
$y<0$ sides. So a bulk test particle will eventually move to the visible
brane at $y=y_c$. This may cause the RS two brane model to be unstable.
Firstly, we note that $y=0$ is an unstable equilibrium position while $y=y_c$
is a stable one. Secondly, it was argued that in sufficiently hard
collisions the SM particles can acquire momentum in the extra dimensions and
escape from the branes\cite{ADD}. As soon as a SM particle was kicked off
the hidden brane at $y=0$ into the bulk, it will be pulled by $F^5$ down to
the visible brane at $y=y_c$. In this way, the distribution of matter on the
two branes can not remain balanced. So we say that the hidden brane is
unstable{\bf .}

In the single brane RS model\cite{RS2}, the $y=$ $y_c$ brane approaches the
AdS horizon. We find that above discussion and conclusion also valid.

It has been noted\cite{CHR} that if the Minkowski metric $\widetilde{\eta }%
_{\alpha \beta }$ in the RS solution (\ref{RSmetr}) is replaced by {\it any}
Ricci flat metric $\widetilde{g}_{\alpha \beta }$ then the Einstein
equations (\ref{RSeq}) are still satisfied\cite{BP}. This enable people to
study {\it any} 4D Einstein's vacuum solutions such as the Schwarzschild one
in the RS scenario. A very interesting work of this kind is discussed in Ref.%
\cite{CHR} where the 5D Schwarzschild solution is called the brane-world
black hole, black string, or black cigar. Here we find that even in the
Ricci-flat case the three equations (\ref{A=5a})-(\ref{A=5c}) still hold.
Therefore, the conclusion is the same that the $y=0$ brane for Ricci-flat
metrics $\widetilde{g}_{\alpha \beta }$ is also unstable.

\section{EQUILIBRIUM EQUATION OF THE BULK FLUID}

To resolve the instability problem, we follow others\cite{BDL}\cite{Kanti4}
to introduce a 5D fluid in the bulk to balance the attraction between the
two branes and to form a hydrodynamical model. This introduction would
generalize the Ricci-flat metric $\widetilde{g}_{\alpha \beta }$ once more
to non Ricci-flat 4D metrics, for which we let 
\begin{equation}
ds^{2}=W^{2}(y)\widetilde{g}_{\alpha \beta }(x^{\mu })dx^{\alpha }dx^{\beta
}+dy^{2}\;,  \label{metr}
\end{equation}%
where $\widetilde{g}_{\alpha \beta }$ is the induced 4D metric. For the bulk
matter, we use anisotropic 5D fluid model and require that the bulk fluid
does not flow along the $y$-direction\cite{Kanti4}, i.e., $u^{5}\equiv
dy/d\tau =0$. That is, we let 
\begin{eqnarray}
T^{AB} &=&\left( 
\begin{array}{ll}
T^{\alpha \beta } & 0 \\ 
0 & P%
\end{array}%
\right) \;,  \nonumber \\
T^{\alpha \beta } &=&(\rho +p)u^{\alpha }u^{\beta }+pg^{\alpha \beta }\;,
\label{T^AB}
\end{eqnarray}%
where $T^{\alpha \beta }$ is of the 4D perfect fluid form with $u^{\alpha
}\equiv dx^{\alpha }/d\tau $. Then by using the fifth equation of the 5D
Bianchi identities $T^{AB};_{B}=0$ we obtain a condition 
\begin{equation}
P^{\prime }=\frac{W^{\prime }}{W}(3p-\rho -4P)\;,  \label{P'}
\end{equation}%
where we have used the relation $g_{\alpha \beta }T^{\alpha \beta }=3p-\rho $
(since $u^{5}=0$) and a prime stands for partial derivative with respect to $%
y$. This condition (\ref{P'}) is a constraint upon the bulk fluid which is
similar to that of star models in astrophysics. Accordingly, we call (\ref%
{P'}) the hydrostatic equilibrium equation of the bulk fluid along the $y$%
-direction.

\section{EXACT 5D BULK SOLUTIONS}

Now the 5D Einstein equations read 
\begin{equation}
R_{AB}-\frac 12g_{AB}R=\frac 1{4M^3}\left\{ T_{AB}-\Lambda g_{AB}-\left[
\lambda _{vis}\delta (y-y_c)+\lambda _{hid}\delta (y)\right] g_{\mu \nu
}\delta _A^\mu \delta _B^\nu \right\} ,  \label{5Deq}
\end{equation}
where $T_{AB}$ takes the form (\ref{T^AB}). To solve equations (\ref{5Deq}),
we firstly use the metric (\ref{metr}) to reduce $R_{AB}-\frac 12g_{AB}R$ to 
\begin{eqnarray}
R_{\alpha \beta }-\frac 12g_{\alpha \beta }R &=&\widetilde{R}_{\alpha \beta
}-\frac 12\widetilde{g}_{\alpha \beta }\widetilde{R}+3\left( WW^{\prime
\prime }+W^{\prime 2}\right) \widetilde{g}_{\alpha \beta }\;,  \nonumber \\
R_{\alpha 5} &=&0\;,  \nonumber \\
R_{55}-\frac 12g_{55}R &=&-\frac 12W^{-2}\widetilde{R}+6W^{-2}W^{\prime
2}\,\;,  \label{G^AB}
\end{eqnarray}
where $\widetilde{R}_{\alpha \beta }$ and $\widetilde{R}$ are made from $%
\widetilde{g}_{\alpha \beta }$. Then, by substituting (\ref{G^AB}) into (\ref%
{5Deq}), we obtain

\begin{equation}
\widetilde{R}_{\alpha \beta }-\frac 12\widetilde{g}_{\alpha \beta }%
\widetilde{R}+3\left( WW^{\prime \prime }+W^{\prime 2}\right) \widetilde{g}%
_{\alpha \beta }=\frac 1{4M^3}\left\{ T_{\alpha \beta }-\left[ \Lambda
+\lambda _{vis}\delta (y-y_c)+\lambda _{hid}\delta (y)\right] g_{\alpha
\beta }\right\} ,\   \label{G^AB1}
\end{equation}
\begin{equation}
-\frac 12\widetilde{R}+6W^{\prime 2}=\frac 1{4M^3}(P-\Lambda )W^2\;.
\label{G^AB2}
\end{equation}
Now we wish to know what kind of exact solutions of Eqs. (\ref{G^AB1}) and (%
\ref{G^AB2}) could fit into the RS two brane model without changing the
boundary conditions (\ref{W(y)}) and (\ref{lambda}). So we suppose that the
``warp'' factor $W(y)$ and the cosmological constants $\Lambda $, $\lambda
_{vis}$, and $\lambda _{hid}$ take the same forms as they do in the RS
solution (\ref{W(y)}) and (\ref{lambda}). Then Eqs. (\ref{G^AB1}) and (\ref%
{G^AB2}) reduce to 
\begin{equation}
\widetilde{R}_{\alpha \beta }-\frac 12\widetilde{g}_{\alpha \beta }%
\widetilde{R}=\frac 1{4M^3}\left[ (\rho +p)u_\alpha u_\beta +pg_{\alpha
\beta }\right] \;,  \label{4eq1}
\end{equation}
\begin{equation}
\widetilde{R}=-\frac 1{2M^3}PW^2\;,  \label{4eq2}
\end{equation}
where we have used (\ref{T^AB}). We see that the left-hand sides of these
two equations are functions of the 4D coordinates $x^\mu $ only. So we wish
somehow to arrange to have the right-hand sides of the two equations also
depend on $x^\mu $ only. To do this, let us define the induced 4D velocity
by $\widetilde{u}^\alpha \equiv dx^\alpha /d\widetilde{\tau }$, where $d%
\widetilde{\tau }^2=-\widetilde{g}_{\alpha \beta }dx^\alpha dx^\beta $. So $%
u^\alpha \equiv dx^\alpha /d\tau =(d\widetilde{\tau }/d\tau )\widetilde{u}%
^\alpha $. Since $u^5=0$ we have 
\begin{equation}
u^\alpha =W^{-1}\widetilde{u}^\alpha \,,\qquad u_\alpha =W\widetilde{u}%
_\alpha \;.  \label{4vel}
\end{equation}
Using this into (\ref{4eq1}) gives 
\begin{equation}
\widetilde{R}_{\alpha \beta }-\frac 12\widetilde{g}_{\alpha \beta }%
\widetilde{R}=\frac 1{4M^3}W^2\left[ (\rho +p)\widetilde{u}_\alpha 
\widetilde{u}_\beta +p\widetilde{g}_{\alpha \beta }\right] \;.
\end{equation}
Note that $\widetilde{u}_\alpha $ and $\widetilde{g}_{\alpha \beta }$ depend
on $x^\mu $ only. So from this equation and equation (\ref{4eq2}) we obtain 
\begin{equation}
\rho =bW^{-2}\widetilde{\rho }\,,\quad \;p=bW^{-2}\widetilde{p}\,,\quad
P=bW^{-2}\widetilde{P}\;,  \label{pP1}
\end{equation}
where $b$ is a constant, $\rho $, $p$ and $P$ satisfy the following
condition 
\begin{equation}
2P=3p-\rho \;,  \label{pP2}
\end{equation}
and $\widetilde{\rho }$, $\widetilde{p}$ and $\widetilde{P}$ are functions
of $x^\mu $ only. Thus, with 
\begin{equation}
b=32M^3\pi G_4\quad ,  \label{b}
\end{equation}
we have successfully brought equations (\ref{4eq1}) to the form of the
standard 4D Einstein's equations with a perfect fluid source: 
\begin{eqnarray}
\widetilde{R}_{\alpha \beta }-\frac 12\widetilde{g}_{\alpha \beta }%
\widetilde{R} &=&8\pi G_4\widetilde{T}_{\alpha \beta }\;,  \nonumber \\
\widetilde{T}_{\alpha \beta } &=&(\widetilde{\rho }+\widetilde{p})\widetilde{%
u}_\alpha \widetilde{u}_\beta +\widetilde{p}\widetilde{g}_{\alpha \beta }\;.
\label{4eq1'}
\end{eqnarray}
Note that equation (\ref{pP2}) plays the same role as the hydrostatic
equilibrium equation (\ref{P'}), and (\ref{P'}) is satisfied automatically.
It is also noticed that some results, such as relations (\ref{pP1}) and (\ref%
{pP2}), are recoveries of previous works\cite{Kanti4}, and is compatible
with global constraints known as the brane world sum rules\cite{GKL}.

Here, we can call $\widetilde{T}_{\alpha \beta }$ in (\ref{4eq1'}) the
effective 4D energy-momentum tensor. Many discussions concerning this kind
of effective or induced energy momentum tensor can be found in the induced
matter theory\cite{Wesson} in which the 4D matter could be a consequence of
the dependence of the 5D metric on the extra dimension. This is also true in
brane models in which if the 5D metric is independent of the extra
dimension, then the brane is void of matter. Detailed discussions for the
relationship between the induced-matter and the brane-world theories can be
found in Ref.\cite{Leon}.

We also wish to emphasis that our derivation for solutions (\ref{pP1}) and (%
\ref{pP2}) is very general. The only restriction is that the 5D bulk
energy-momentum tensor $T^{AB}$ should take a fluid form (\ref{T^AB}). Since
(\ref{4eq1'}) are just the 4D Einstein equations, we can conclude that {\it %
any} known 4D exact solution, which has a perfect fluid as source, can be
embedded in 4D hypersurfaces of the bulk to generate a 5D exact solution of
the 5D equations (\ref{5Deq}), with 5D metric as in (\ref{metr}), the
``warp'' factor $W(y)$ and the cosmological constants $\Lambda $, $\lambda
_{vis}$, $\lambda _{hid}$ as in (\ref{W(y)}) and (\ref{lambda}), and $%
\widetilde{\rho }$, $\widetilde{p}$, $\widetilde{P}$ satisfy relations (\ref%
{pP1}) and (\ref{pP2}).

\section{EMBEDDING OF 4D EINSTEIN SOLUTIONS}

It is well known that most 4D exact solutions of general relativity have
used a perfect fluid as source, such as the standard FRW cosmological
solutions and various exterior and interior solutions for various rotating
and non-rotating neutral stars. Using the relations obtained in section 4,
all these solutions can easily be embedded in the RS model to form 5D exact
solutions without changing the RS boundaries. For example, the 5D FRW
cosmological solutions are 
\begin{equation}
ds^2=e^{-2k\left| y\right| }\left[ -dt^2+a^2(t)\left( \frac{dr^2}{%
1-k^{\prime }r^2}+r^2d\Omega ^2\right) \right] +dy^2\;,  \label{CosS1}
\end{equation}
where $k^{\prime }$ is the 3D curvature index ($k^{\prime }=\pm 1,0$), $%
d\Omega ^2\equiv d\theta ^2+\sin ^2\theta d\varphi ^2$, and 
\begin{eqnarray}
\left( \frac{da}{dt}\right) ^2+k^{\prime } &=&\frac{8\pi G_4}3\widetilde{%
\rho }a^2\quad ,  \nonumber \\
a^3\frac{d\widetilde{p}}{dt} &=&\frac d{dt}\left[ \left( \widetilde{\rho }+%
\widetilde{p}\right) a^3\right] \quad ,  \nonumber \\
\rho &=&\left( G_4/G_5\right) e^{2k\left| y\right| }\widetilde{\rho }%
(t)\;,\qquad  \nonumber \\
p &=&\left( G_4/G_5\right) e^{2k\left| y\right| }\widetilde{p}(t)\;, 
\nonumber \\
2P &=&\left( G_4/G_5\right) e^{2k\left| y\right| }\left[ 3\widetilde{p}(t)-%
\widetilde{\rho }(t)\right] \;,  \label{CosS2}
\end{eqnarray}
where $8\pi G_5=(4M^3)^{-1}$. From these we see that $\rho ,$ $p$ and $P$
increase exponentially when $y$ tends from the hidden brane at $y=0$ to the
visible brane at $y=y_c$.

As a second example, we write down the Schwarzschild-AdS$_5$ solution in the
following: 
\begin{equation}
ds^2=e^{-2k\left| y\right| }\left[ -U(r)dt^2+U(r)^{-1}dr^2+r^2d\Omega ^2%
\right] +dy^2\;,  \label{S-AdS1}
\end{equation}
where

\begin{equation}
U(r)=1-\frac{2G_4M}r+\frac 13\widetilde{\lambda }r^2\;,  \label{S-AdS2}
\end{equation}
and 
\begin{eqnarray}
-\widetilde{\rho } &=&\widetilde{p}=\frac 1{8\pi G_4}\widetilde{\lambda }\;,
\nonumber \\
-\rho &=&p=\frac 1{8\pi G_5}e^{2k\left| y\right| }\widetilde{\lambda }\;, 
\nonumber \\
P &=&\frac 1{4\pi G_5}e^{2k\left| y\right| }\widetilde{\lambda }\;.
\label{S-AdS3}
\end{eqnarray}
So in the vicinity of the visible brane, the magnitude of the 5D densities $%
\rho $, $p$ and $P$ are much larger than those in the vicinity of the hidden
brane.

By using known 4D exact solutions, more 5D exact solutions can be obtained
easily in this way. We can show that the hydrostatic equilibrium equation (%
\ref{P'}) is satisfied by all these solutions. So all these solutions are
equilibrium states of the RS model.

\section{COMPACTIFICATION SIZE OF THE FIFTH DIMENSION}

By introducing a scalar field in the bulk, Goldberger and Wise\cite{GW}
proposed a dynamics to stabilize the size of the extra dimension. The
mechanism was to integrate the scalar field action over the fifth dimension
to yield an effective 4D potential. Then it was found that this potential
has a minimum which yields a compactification radius without fine tuning of
parameters. In our case, there is no scalar field available in the bulk. If
one still wish to stabilize the extra dimension, one may need look for
another quantity to minimum.

For simplicity, let us consider the 5D Schwarzschild-AdS$_5$ solution (\ref%
{S-AdS1})-(\ref{S-AdS3}). For this solution the 5D scalar curvature $R$ can
be calculated by using (\ref{5Deq}), (\ref{T^AB}), (\ref{S-AdS3}) and the
relation $8\pi G_5=(4M^3)^{-1}$as 
\begin{equation}
R=-\frac 1{6M^3}\left\{ 24M^3\widetilde{\lambda }e^{2k\left| y\right|
}-5\Lambda -4\left[ \lambda _{vis}\delta (y-y_c)+\lambda _{hid}\delta (y)%
\right] \right\} .  \label{R}
\end{equation}
Now we consider the geometrical part of the 5D action 
\begin{equation}
S_{geo}=\int d^4x\int_{-y_c}^{y_c}2M^3\sqrt{-g}Rdy\quad .  \label{Sgeo}
\end{equation}
Substituting (\ref{R}) into this equation and integrating over the fifth
dimension, we obtain 
\begin{equation}
S_{geo}=-\frac 13\int \sqrt{-\widetilde{g}}d^4x\left[ \frac{24}kM^3%
\widetilde{\lambda }\left( 1-e^{-2ky_c}\right) -\frac 5{2k}\Lambda \left(
1-e^{-4ky_c}\right) -4\lambda _{vis}e^{-4ky_c}-4\lambda _{hid}\right] .
\label{Sgeo2}
\end{equation}
Denote the expression inside the square bracket of this equation as $K$, and
use (\ref{lambda}) to eliminate $\Lambda $, $\lambda _{vis}$, and $\lambda
_{hid}$ in $K$, we find 
\begin{equation}
K\equiv -6M^3\int_{-y_c}^{y_c}W^4Rdy=\frac{24}kM^3\widetilde{\lambda }\left(
1-e^{-2ky_c}\right) -36M^3k\left( 1-e^{-4ky_c}\right) \;.  \label{K}
\end{equation}
This $K$ can be interpreted as an effective 4D curvature. Interestingly we
find that this $K$ has a minimum at 
\begin{equation}
e^{-2ky_c}=\frac{\widetilde{\lambda }}{3k^2}\quad ,  \label{Yc}
\end{equation}
at which 
\begin{equation}
\frac{\partial ^2K}{\partial y_c^2}=\frac{32}kM^3\widetilde{\lambda }^2>0\;.
\label{K''b}
\end{equation}
Therefore we see that the relation (\ref{Yc}) may provide us with a possible
compactification size $y_c$ for the fifth dimension.

Note that the relation (\ref{Yc}) requires $\widetilde{\lambda }$ being
positive. Be aware that if $\widetilde{\lambda }$ is negative, then the 5D
Schwarzschild-AdS$_5$ solution (\ref{S-AdS1})-(\ref{S-AdS3}) becomes the 5D
Schwarzschild-dS$_5$ solution. So the effective 4D curvature $K$ of the 5D
Schwarzschild-dS$_5$ solution does not have a minimum.

From (\ref{S-AdS2}) it is reasonable to expect $\widetilde{\lambda }%
y_{c}^{2}\ll 1$. For instance, if $\widetilde{\lambda }y_{c}^{2}=10^{-31}$,
then $ky_{c}\simeq 40$. We find that this value of $y_{c}$ meets the
requirement from the hierarchy problem\cite{RS1}.

\section{CONCLUSION}

In this paper we have studied the gravitational force field in the bulk of
the RS two brane model by using the 5D geodesic equations. This force may
cause the hidden brane to be unstable. To balance this force we have
introduced a 5D fluid in the bulk with it's 4D part being a perfect fluid.
Thus a hydrostatic equilibrium equation for the bulk fluid is derived.
Meanwhile, a class of exact bulk solutions is obtained. In 4D hypersurfaces
these solutions turn out to be exactly the same as the 4D Einstein equations
with a perfect fluid source. Therefore, one can obtain exact 5D bulk
solutions by simply embedding a suitable 4D solution in the bulk. Then we
have discussed the stabilization size of the extra dimension. Further
investigation is needed.

\bigskip \noindent {\bf ACKNOWLEDGMENTS}

This work was supported by the National Natural Science Foundation of China
under grant 19975007.

\end{document}